\documentclass[prb,twocolumn,superscriptaddress]{revtex4}

\usepackage{epsfig}
\usepackage{amsmath}
\usepackage{graphicx}

\begin{document}

\title{From multiferroics to cosmology: Scaling behaviour and beyond in the hexagonal manganites}

\author{S. M. Griffin}
\affiliation{Department of Materials, ETH Zurich, Wolfgang-Pauli-Strasse 10, CH-8093 Zurich, Switzerland}
\author{M. Lilienblum}
\affiliation{Department of Materials, ETH Zurich, Wolfgang-Pauli-Strasse 10, CH-8093 Zurich, Switzerland}
\author{K. Delaney}
\affiliation{Materials Research Laboratory, University of California, Santa Barbara, California 93106, USA}
\author{Y. Kumagai}
\affiliation{Department of Materials, ETH Zurich, Wolfgang-Pauli-Strasse 10, CH-8093 Zurich, Switzerland}
\author{M. Fiebig}
\affiliation{Department of Materials, ETH Zurich, Wolfgang-Pauli-Strasse 10, CH-8093 Zurich, Switzerland}
\author{N. A. Spaldin}
\affiliation{Department of Materials, ETH Zurich, Wolfgang-Pauli-Strasse 10, CH-8093 Zurich, Switzerland}

\date{\today}

\begin{abstract}
We show that the improper ferroelectric phase transition in the multiferroic hexagonal manganites 
displays the same symmetry-breaking characteristics as those proposed in early-universe theories. 
We present an analysis of the Kibble-Zurek
theory of topological defect formation applied to the hexagonal manganites, discuss the conditions
determining the range of cooling rates in which Kibble-Zurek behavior is expected, and show that
recent literature data\cite{Chae_et_al:2012} are consistent with our predictions. 
We explore experimentally for
the first time to our knowledge the cross-over out of the Kibble-Zurek
regime\cite{Yates/Zurek:1998} and find a surprising ``anti-Kibble-Zurek'' behavior. 
\end{abstract}

\maketitle

A key open question in cosmology is whether the vacuum contains topological defects
such as cosmic strings \cite{Hindmarsh/Ringeval/Suyama:2010, CosmicStrings, Vachaspati:2009,
Bevis/Hindmarsh/Kunz/Urrestilla:2008, Jeannerot/Rocher/Sakellariadou:2003}. 
The formation of topological defects during phase transitions in the early universe -- believed in
grand unified theories to occur within around 10$^{-34}$ s after the Big Bang -- was proposed by
Kibble\cite{Kibble:1976}, who derived the symmetry requirements for formation of topological
defects in a physical manifold. In systems where such defects are symmetry allowed, their density
can be estimated using the Zurek scenario\cite{Zurek:1985}, which uses causality arguments to
develop scaling laws for the density of defects formed as a function of the rate of quenching
across the phase transition. The resulting combination of symmetry requirements and scaling laws
is termed the Kibble-Zurek mechanism, and in principle should describe a phase transition in any
system with the required symmetry properties, provided that other effects do not dominate the
kinetics of topological defect formation.

Attempts to demonstrate Kibble-Zurek scaling in condensed-matter systems to date have proved 
challenging, however, and the ``ideal Kibble-Zurek system'' has remained elusive.
Zurek's original paper\cite{Zurek:1985} discussed the analogue between cosmic strings and the
vortex cores formed in a quench-induced phase transition from normal-state to superfluid $^{4}$He.
However the corresponding experiment\cite{Hendry:1994, Dodd:1998} yielded large deviations from
the predicted behavior, probably because of thermal effects\cite{Rivers:2000}. In $^{3}$He
the symmetry breaking is closer to that postulated for the early universe\cite{Bauerle_et_al:1996,
Ruutu:1996}, but the density of topological defects can only be inferred indirectly and
many assumptions must be made to compare with predictions. 
In superconducting Nb rings the density of vortex cores in the superconducting current 
led to a different scaling exponent than that predicted by the
Kibble-Zurek mechanism --  again, experimental artifacts were held
responsible\cite{Monaco_et_al:2009} -- and 
no flux-line formation at all was observed in experiments
on high-$T_{C}$ superconducting thin films\cite{Carmi/Polturak:1999}. Bose-Einstein
condensates could in principle provide a suitable system but are so far subject to experimental
limitations\cite{Weiler_et_al:2008}. 
Perhaps the best candidates to date have been liquid crystals, in which the
diffraction pattern formation in nonlinear-optic experiments has been shown to exhibit
power-law scaling \cite{Ducci_et_al:1999} and promising studies 
of defect dynamics have been performed\cite{Chuang_et_al:1991}, but 
strong interactions between the defects continue to cause difficulties.

Here we propose the multiferroic hexagonal magnanites, \textit{R}MnO$_{3}$, \textit{R} = Sc, Y, Dy
-- Lu, as a model system for testing the Kibble-Zurek mechanism. The hexagonal manganites have
attracted interest because of their unusual geometrically-driven improper ferroelectricity, which
allows for the simultaneous occurrence of magnetic ordering\cite{vanAken_et_al:2004,
Fennie/Rabe_YMO:2005}, as well as unusual
couplings\cite{Fiebig_et_al:2002,Lottermoser_et_al:2004,Choi_et_al:2010} and
functionalities\cite{Meier_et_al:2012} at their domain walls. In this work we show that the
unusual nature of the improper geometric ferroelectric phase transition also sets both the correct
symmetry conditions for Kibble-Zurek behavior to manifest, as well as the physical properties for
readily detecting it. In addition, the relevant time-, temperature- and length-scales fall into
a range that allows exploration of the Kibble-Zurek regime, as well as the crossover out of it.

\section{Symmetry and physical properties of $R$MnO$_3$}

\begin{figure}
 \centering
\includegraphics[width=1.0\linewidth]{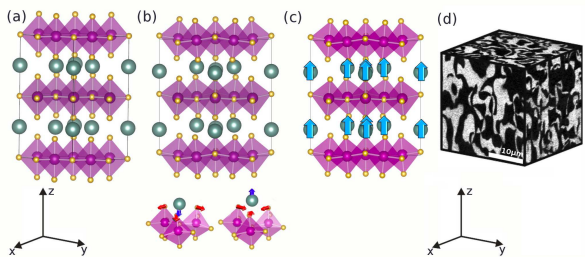}
 \caption{(a) High-symmetry $P6_3/mmc$ structure of $R$MnO$_{3}$ before the onset of
trimerization. (b) Action of the $K_3$ trimerization mode on the $R$ ions and MnO$_{5}$ trigonal
bipyramids. The insets below the main structure are to emphasize that outward trimerization 
results in a downward shift of the corresponding $R$ ion, whereas inward trimerization results 
in an upward shift.
(c) The subsequent additional displacements of the $R$ ions (blue arrows) in the $\Gamma_2^-$ mode provide the
ferroelectricity. Note that once the orientation of the trimerization mode is set, the spontaneous
polarization can emerge in only one direction. (d) Typical domain structure measured using piezoforce
microscopy. The black and white regions correspond to opposite orientations of the ferroelectric
polarization along the z axis. Note that the domain structure is isotropic, in spite of the layered
crystal structure.}
 \label{Fig1}
\end{figure}

First we describe the properties of $R$MnO$_3$ that are relevant for testing the Kibble-Zurek
mechanism, particularly the symmetry properties of the phase transition. The structure of
$R$MnO$_3$ consists of planes of MnO$_5$ trigonal bipyramids separated by planes of $R$ ions which
form a hexagonal mesh (Fig.\ 1(a))\cite{Yakel_et_al:1963}. In the high-temperature
paraelectric phase, the space group is centrosymmetric $P6_3/mmc$. At $T_C \sim 1400$~K (the exact
value depends on the $R$ ion) a spontaneous symmetry breaking occurs, with the condensation of 
primarily two phonon modes with distinct
irreducible representations of the high-symmetry 
structure\cite{Lonkai_et_al:2004,Fennie/Rabe_YMO:2005}. First, a mode of $K_3$ symmetry, which involves a
trimerizing tilt of the trigonal bipyramids and is the primary order parameter
(Fig.\ 1(b)). Since the $K_3$ mode can condense about three different origins, and the
tilt can be in the ``in'' or ``out'' direction, six trimerization domains are formed; these have
been shown using high resolution transmission electron microscopy to meet at vortex
cores\cite{Choi_et_al:2010}. Importantly (and unusually), while this mode lowers the symmetry to
that of a polar space group it carries no net polarization, as any net local polarity vanishes
macroscopically due to the non-zero mode wave vector. A secondary mode of $\Gamma_2^-$
symmetry (referring to the parent space group), which does not further lower the symmetry, provides 
the ferroelectric polarization
(Fig.\ 1(c)). While the orientation of this secondary ferroelectric polarization is set by
the in- or out- tilt of the $K_3$ mode and does not result in additional domains, it is essential
for our experiments as it allows the straightforward imaging of the domain structure
using piezoresponse force microscopy (PFM). Indeed PFM measurements reveal that domains of
alternating polarization are locked to the trimerization domains around vortex 
cores\cite{Choi_et_al:2010, Jungk_et_al:2010} yielding appealing
six-fold patterns (Fig.\ 1(d)). Electric-field poling experiments have shown that the vortex cores are
protected in the sense that they cannot be annihilated or driven out of the system
by an electric field\cite{Choi_et_al:2010, Jungk_et_al:2010}. Surprisingly, the domain structure
and density of these topological defects when viewed from the side of the sample is similar to that
viewed from the top in spite of the layered crystal structure and uniaxial
ferroelectricity\cite{Jungk_et_al:2010} (Fig.\ 1(d)). This absence of anisotropy in the domain 
structure allows for straightforward determination of the defect densities from two-dimensional 
top-view scans of their areal density, rather than requiring a complex three-dimensional analysis.

First-principles calculations\cite{Fennie/Rabe_YMO:2005} and Landau Theory
analysis\cite{Artyukhin_et_al:2012} have shown that for small magnitudes of the trimerizing $K_3$
mode, the polar mode appears only as a third-order term, and so the magnitude of the ferroelectric
polarization just below $T_C$ is vanishingly small. This is important for two reasons:
First, the formation of the domain structure at $T_C$ is not influenced by the system's attempts
to minimize the depolarizing field from the ferroelectric polarization. Strong evidence for this
is given by the large numbers of electrostatically unfavorable head-to-head and tail-to-tail
domain walls that form in $R$MnO$_3$, but rarely occur in conventional
ferroelectrics\cite{Meier_et_al:2012}. Second, first-principles calculations show that the energy
lowering provided by the condensation of the $K_3$ mode is independent of the angle of the tilt
until the polar mode subsequently develops\footnote{Note that the Landau free energy derived in Ref.
\cite{Artyukhin_et_al:2012} has a branch cut in the trimerization angle that gives a signature of
topological protection if the angle rotates through the branch cut when traced round the vortex core.}
\cite{Artyukhin_et_al:2012}. This means that the
potential below the phase transition temperature is given by the continuous ``Mexican hat'' form
(Fig.\ 2).  The discrete nature of the lattice does not manifest until lower temperatures 
when the domain structure is already determined.
As a result we can use the mathematics of
continuous symmetries which are usually assumed in the Kibble-Zurek mechanism. In this language,
the full rotational symmetry is broken when the polyhedra tilt in the $2\pi$ range of angles,
resulting in a $U(1)$ vacuum.

\begin{figure}
\centering
 \includegraphics[width=0.8\linewidth]{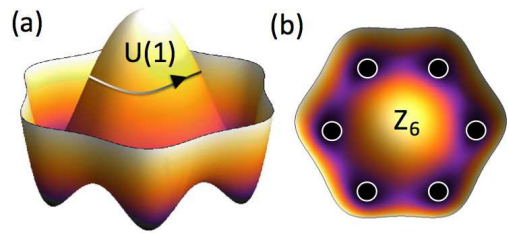}
\caption{Mexican-hat potential energy surface of the hexagonal manganites. At high energy 
(the peak of the hat) the energy is independent of the angle of trimerization, and the system 
has $U(1)$ symmetry. At lower energy (in the brim of the hat), six of the trimerization angles
become favorable (white circles), and the symmetry reduces to the sixfold discrete symmetry 
described by $Z_{6}$.}
 \label{Fig2}
\end{figure}

For larger magnitudes of the $K_3$ mode, obtained upon temperature decrease, a cross-over to
linear coupling with the polar $\Gamma_2^-$ mode occurs and the polarization becomes measurably
large. This lifts the degeneracy with angle of the $K_3$ mode and fixes the polyhedra into
discrete tilt angles of $0, 2\pi/3$ or $4\pi/3$, described by $Z_{3}$ symmetry. The additional
degeneracy provided by the direction (``in'' or ``out'') of the polyhedral tilting gives an
additional $Z_{2}$ symmetry reduction, resulting in $Z_{2}\times Z_{3} = Z_{6}$. It is an open
experimental question whether the onset of the $\Gamma_2^-$ mode, which is observed $\sim 300$ K
below $T_C$ is an ``emergence'' or an additional isosymmetric phase 
transition\cite{Lonkai_et_al:2004,Fennie/Rabe_YMO:2005}.

\section{Kibble-Zurek scenario for \textit{R}MnO$_3$}

In this section we first show that the symmetry of $R$MnO$_3$ results in topologically protected vortex
cores as described by the Kibble mechanism. We then analyze the vortex cores using the Zurek
scenario to determine the density of topological defects that should be produced as a function of
the cooling rate through the phase transition. We use first-principles density-functional theory
to evaluate the relevant parameters, and show that our predictions are in excellent agreement
with literature data. 

\subsection{Kibble mechanism and the formation of topological defects.}

The requirements for the formation of topological defects at a phase transition within the Kibble
mechanism\cite{Kibble:1976} are (i) a spontaneous symmetry breaking and (ii) a change in symmetry
across the phase transition that corresponds to a non-trivial homotopy group. The
trimerization transition in $R$MnO$_3$ clearly fulfils the first condition; next we show that it
also fulfils the second.

As discussed earlier, in the temperature range just below the phase transition, $R$MnO$_3$ exhibits
a continuous symmetry; this allows us to use the methods and results of homotopy theory -- which are 
developed for continuous symmetry groups -- to assess the topology of $R$MnO$_3$. It is established 
within homotopy theory that the symmetry characteristics of the order parameter, in our case $U(1)$, 
can be used to assess the topological characteristics of a phase transition. To make the 
assessment, the order parameter symmetry is first mapped onto an $n$-dimensional sphere. In the
case of the $U(1)$ symmetry, this is a one-dimensional circle, $S^1$. Next a function called the 
homotopy group, $\pi_{k}$, which describes the topological nature of the order parameter symmetry
is defined. If $\pi_k$ differs from the identity, then it is non-trivial and topological defects 
are formed. It has been known since the 1960's\cite{Kervaire:1963} that $\pi_k(S^{1})$ is indeed
non-trivial and in fact produces one-dimensional topological singularities, called strings or
vortex cores\cite{Kibble:2000}. Therefore the vortex cores in $R$MnO$_3$ are mathematically
topologically protected, in concordance with their physical topological protection -- their
resistance to annihilation by an electric field -- that we discussed
earlier\cite{Choi_et_al:2010,Jungk_et_al:2010}. We note also that within the Kibble mechanism the
topological defects are remnants of the parent phase trapped within the lower symmetry phase. For
$R$MnO$_3$ this implies that the high-symmetry paraelectric phase is preserved at the meeting
point of the six domains defining a vortex core.

\subsection{Zurek scenario for $R$MnO$_{3}$.}

Within the Zurek scenario, the density of topological defects formed during a spontaneous symmetry
breaking phase transition described by the Kibble mechanism follows a power-law dependence on the
rate at which the transition is crossed\cite{Zurek:1985,PhysRevLett.85.4660}. In this section we 
first relate the material properties
of $R$MnO$_3$ to the parameters in the Zurek scenario. We then evaluate their magnitudes to
predict quantitatively the temperature dependence of the defect formation within the Kibble-Zurek
mechanism.

Zurek's approach relies on the notion of competing timescales: The first relevant timescale
is the time it takes for one region of the system to communicate its choice of vacuum state 
with another. This time becomes divergently long towards the critical temperature as the correlation length 
diverges, a phenomenon termed ``critical slowing down''. The second is the quench time
$\tau_q$ that the system spends cooling through the phase transition. 
The size of the domains is set at the temperature
$T_C + \Delta T_f$ where the communication distance across which information can be
transferred during the progressing phase transition becomes equal to the correlation length
$\xi(T)$ (Fig.\ 3). While the correlation $\xi(T)$ continues to diverge as the temperature further 
approaches $T_C$, the communication length remains unchanged, and the
system is unable to adapt to its increase. 
As consequence a ``freeze-out'' occurs in the temperature
interval between $T_C + \Delta T_f$ and $T_{C}-\Delta T_f$: The size of the correlated regions
is unable to 
increase and so the domain size is fixed at the value $\xi(T_C + \Delta T_f)$ (Fig.\ 3). For fast
cooling through the transition, the distance over which information can be transferred during the
transition is small, and becomes equal to the correlation length at small values of $\xi(T)$.
Therefore freeze-out occurs when the domain size is small (and consequently the number of
topological defects is large). In contrast, for slow cooling, the distance for information
transfer is large, and does not become equal to $\xi(T)$ until close to the phase transition
temperature, where $\xi(T)$ is large. In this case large domains, with fewer topological defects,
form.

\begin{figure}
\centering
 \includegraphics[width=\linewidth]{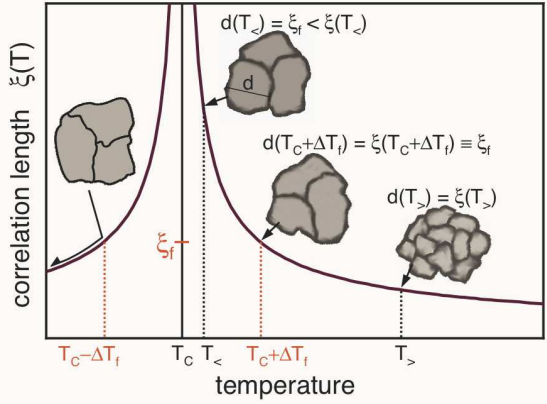}
\caption{Domain formation in the Kibble-Zurek scenario. Above $T_C$ fluctuating regions of 
lateral extension $d$ with uniform orientation of the emerging order parameter occur (fuzzy
patches). At high temperature ($T_> > T_C + \Delta T_f$) the size of the correlated regions
is determined by the
correlation length (purple curve). At temperature $T = T_C + \Delta T_f$ a freeze-out of the 
lateral extension $d$ begins. Below the freeze-out temperature, 
Fluctuations continue to occur but the lateral extension of the fluctuating regions can no
longer follow
the diverging correlation length while the system continues to cool. The size of the
fluctuation regions in this range, for example at $T_<$ remains set by the correlation 
length at the freeze-out
temperature, $\xi(T_C + \Delta T_f)$, and corresponds to 
the ``communication length'' which is the distance that information can
propagate during the time in which the system cools from $T_C + \Delta T_f $ to 
$T_C- \Delta T_f$.
Below $T_C- \Delta T_f$ stable domains of the lateral
extension $\xi(T_C + \Delta T_f)$ form (areas confined by black lines).}
 \label{Fig3}
\end{figure}

Here we summarize the derivation of the density of topological defects as a function of
quench rate through the phase transition within the Kibble-Zurek mechanism. For a detailed
derivation we recommend particularly Ref. \cite{Kibble:2003}. First, we use
critical scaling: As the system approaches $T_{C}$, the correlation length, $\xi$, and relaxation time, 
$\tau$, diverge as
\begin{eqnarray*}
 \xi(T) &=& \xi_{0}\left|1-\frac{T}{T_C}\right|^{-\nu} \\
 \tau(T) &=& \tau_{0}\left|1-\frac{T}{T_C}\right|^{-\mu} 
\end{eqnarray*}
where $\xi_{0}$ is the zero-temperature correlation length and $\tau_{0}$ is the zero-temperature 
time, which is equal to $\xi_0$ divided by the speed of information transfer in the system. 
Both $\xi_0$ and $\tau_0$ are system-dependent quantities. 
$\nu$ and $\mu$ are critical exponents that are determined by the
universality class of the phase transition, that is its general behaviour as determined by the
symmetry properties of the phase transition, irrespective of the material properties of the
specific system. 

Assuming that the temperature varies linearly with time near the phase transition, and taking $t=0$
at $T=T_C$, it is clear that $T = T_C + t \times r_q$, where $r_q$ is the cooling rate. Rearranging this expression yields
\begin{equation*}
\left| 1- \frac{T}{T_C} \right| = \frac{r_q}{T_C}t = \frac{t}{\tau_{q}} 
\end{equation*}
where $\tau_q = \frac{T_C}{r_q}$ divided by the cooling rate, is called the ``quench time''.

The speed at which information is transferred in the material is then given by the characteristic velocity,
\begin{equation*}
 c(T) = \frac{\xi(T)}{\tau(T)} = \frac{\xi_{0}}{\tau_{0}}\left|1-\frac{T}{T_C}\right|^{\mu-\nu}
\end{equation*}
and the corresponding distance over which information can propagate in time $t$ is
\begin{eqnarray*}
\int_{0}^{t} c \left(T(t') \right) dt' 
& = & \frac{\xi_0}{\tau_0} \int_0^t \left(\frac{t'}{\tau_q}\right)^{\mu-\nu} dt'\\
& = & \frac{1}{1+\mu-\nu}\frac{\xi_{0}}{\tau_{0}}\tau_q\left|1-\frac{T}{T_C}\right|^{1+\mu-\nu}
\end{eqnarray*}
where in the last step we have substituted $t = \left|1-\frac{T}{T_C}\right| \tau_q$.

Equating the distance over which information can propagate to the correlation length yields an
expression for the freeze-in temperature, $T_f$: 
\begin{equation*}
\frac{1}{1+\mu-\nu}\frac{\xi_{0}}{\tau_{0}}\tau_q\left|1-\frac{T_f}{T_C}\right|^{1+\mu-\nu}
 =  \xi_0 \left|1-\frac{T_f}{T_C}\right|^{-\nu} \\
\end{equation*}
so
\begin{equation*}
 \left|1-\frac{T_f}{T_C}\right|= \left( (1+\mu-\nu) \frac{\tau_{0}}{\tau_{q}} \right)^{\frac{1}{1+\mu}}
\end{equation*}
At temperature $T_f$, the domain sizes are frozen in with a characteristic length-scale given by the information 
propagation distance (which is equal to the correlation length) at the freeze-in temperature: 
\begin{equation*}
\xi_f = \xi_{0} \left(1+\mu-\nu\right)^{\frac{\nu}{1+\mu}} \left( \frac{\tau_{q}}{\tau_{0}} \right)^{\frac{\nu}{1+\mu}}
\end{equation*}
Formally, this corresponds to the length-scale of the vortex ``strings'', that is the three-dimensional
continuation of the vortex cores through the bulk of the sample. The number of vortex intersections
per unit area, $n$, is then approximately equal to the length of vortices per unit volume, $\frac{1}{\xi_f^2}$, giving
\begin{equation*}
 n \approx \frac{1}{\xi_{0}^{2}} \left( \frac{\tau_{0}}{\tau_{q}} \right)^{\frac{2 \nu}{1+\mu}}  \quad .
\label{noftau}
\end{equation*}

To apply the linear Zurek scaling law given in Eq.~(\ref{noftau}) to the hexagonal manganites we
next identify the relevant time- and length-scales in the system, and evaluate their magnitudes.
Our electronic structure calculations were performed using density functional theory
within the local density plus Hubbard-$U$ approximation following the Liechtenstein
approach\cite{Lichtenstein_et_al:1995} with the double-counting corrections treated in the fully
localized limit. Following previous literature studies\cite{Fennie/Rabe_YMO:2005}, we set the
LDA+$U$ parameters on the Mn $3d$ orbitals to $U=8$ and $J=0.88$~eV respectively and enforced an
A-type antiferromagnetic ordering. We used the projector-augmented wave method for core-valence
partitioning\cite{Bloechl:1994}, which significantly reduces the required plane-wave energy cutoff,
and carefully tested the convergence of plane-wave cutoff and k-point sampling.

The zero-temperature correlation length, $\xi_0$, is usually equated with the zero-temperature
domain wall width in ferroelectrics. In order to extract this value we performed density
functional calculations within the LDA$+U$ method using the {\sc VASP}
code~\cite{Kresse/Hafner:1993,Kresse/Furthmuller:1996}. We constructed supercells containing two
180$^{\circ}$
domain walls and in turn 120, 180, 240 and 300 atoms. We fixed the lattice constants of the
supercells to those of the relaxed single-domain unit cell, and then optimized the internal
positions until the forces acting on all atoms converged to less than 0.01~eV/\AA\ respectively.
We initialized a different trimerization phase and ferroelectric orientation within adjacent
domains, and subsequently perfomed full relaxations on the structures; in all cases the system
remained in the metastable multi-domain state. For all supercell sizes, we found that the
structural phase defined by either the tilt of the MnO bipyramids or the direction of off-centring
of the Y ions changes abruptly at the domain walls, indicating an effective domain wall width
close to zero. This is 
consistent with a recent experimental electron microscopy study\cite{Zhang_et_al:2012}, and
sets an upper limit on $\xi_0$ of  $\sim 1$~\AA. 

To calculate the characteristic timescale of the 
system, $\tau_{0} = \frac{\xi_0}{s}$, we require $s$, which is the speed at which the system 
communicates the lattice
distortion as it passes through the phase transition. For solid-state systems, $s$ is given by the
speed of sound. 
To calculate the speed of sound at zero kelvin we used the \verb!ABINIT!\footnote{The ABINIT code is a common
project of the Universit\'e Catholique de Louvain, Corning Incorporated, and other contributors
(URL http://www.abinit.org).} software package\cite{Gonze_et_al:2002,Gonze_et_al:2009} to optimize the structure of
a 10-atom unit cell, then calculated the full phonon band structure using frozen-phonon
techniques. A supercell was constructed with doubling and trebling in each of the directions
required to sample the first Brillouin zone, then symmetry-distinct displacements were made to
construct the full matrix of interatomic force constants. The dynamical matrix was diagonalized
along each of the high symmetry lines shown in the phonon band structure using Fourier
interpolation.\cite{phonopy, Parlinski_et_al:1997} The speed of sound was then extracted from the
calculated phonon band structure by fitting the acoustic branch with a polynomial, then evaluating
the group velocity
\[ v_g = \left.\frac{\partial \omega}{\partial \vec{k}}\right|_{\vec{k}=0}\] \quad . 
The relevant velocity for our Kibble-Zurek fit is the doubly degenerate branch with the
atoms displacing in plane and the wavevector propagating in plane. For this branch we obtained
$v_g = 640$ m.s$^{-1}$.

Finally, for comparison with quenching experiments we also 
need the Curie temperature, $T_C$, which
relates $\tau_q$ to the cooling rate, $r_q$ through $r_q = \frac{T_C}{\tau_Q}$. This is known experimentally
to be $\sim1400$ K, with the exact number depending on the $R$ ion. 

Next we extract the critical exponents by identifying that, as in the case of the superfluid
transition in $^4$He, the $R$MnO$_3$ transition belongs to the universality class denoted as the
3D XY
model. The values of the critical exponents for this universality class have been calculated using
Monte Carlo simulations\cite{Campostrini_et_al:2006} to be $\nu = 0.6717$ and $\mu = 1.3132$,
giving a Kibble-Zurek scaling exponent of $\frac{2\nu}{1+\mu}\approx 0.29$.
Taking the values introduced so far, with our upper limit for $\xi_0$, 
we find that domains of $5\mu m$ width should be formed for
a quench time of $\tau_q \sim 40$~min (corresponding to a cooling rate of $\sim 0.5$~K/s), and 
domains of 40~$\mu$m in around one month (cooling rate $\sim 1.5$~K/hour. These are readily
experimentally accessible cooling rates. 

In Fig. 5(d) we compare our predicted scaling behavior with 
recently reported vortex densities measured as a function of cooling
rate in ErMnO$_3$ (red circles)\cite{Chae_et_al:2012}. 
The agreement in scaling behavior between the experiment and the Kibble-Zurek prediction
is excellent. Note in particular that the scaling exponent matches exactly the theoretical
prediction; this is the first time to our knowledge that such good agreement has been
obtained. We obtain the best match with a value of zero-temperature correlation length 
of $0.06$~\AA{} (solid red line), consistent with the approximately zero domain wall width  
obtained in our density functional calculations. 
Hence, $R$MnO$_3$ is the first system to show model Kibble-Zurek behavior
in the laboratory. 

We thus find a unique situation in
$R$MnO$_3$. Topologically, it is a model system for the experimental verification of the
Kibble-Zurek mechanism. In the temperature range where the Kibble-Zurek mechanism is expected to
govern the formation of domains and the distribution of vortex-core singularities, the unwanted
ferroelectric polarization which could influence domain formation is effectively absent. However,
at room temperature the coupling of the distortive order parameter to the now finite ferroelectric
polarization allows straightforward imaging of the topology and vortices via spatially
resolved measurements of the ferroelectric domain structure. And finally, a distinguishable range
of domain sizes and hence defect densities is obtained for an experimentally accessible range of
cooling rates through the primary distortive phase transition at $T_C$.

\section{Beyond the Kibble-Zurek limit}

The Kibble-Zurek mechanism applies only to the regime in which the system has time to respond
adiabatically to the cooling until the freeze-out temperature, $T_C + \Delta T_f$ is reached. 
For faster quenching, it is expected that the
Kibble-Zurek mechanism should break down and be replaced by a dynamics that is largely unknown. 
The unique combination of properties in $R$MnO$_3$ allows us to continue to increase
the cooling rate beyond those explored in Ref.~\cite{Chae_et_al:2012} to  investigate
the evolution out of the Kibble-Zurek regime; we explore this range 
next. 

For our experiment we chose YMnO$_3$ rather than
the ErMnO$_3$ that was used in Ref.~\cite{Chae_et_al:2012} because of the greater thickness
of our YMnO$_3$ samples. 
The experimental procedure applied to the
flux-grown, c-oriented single crystal YMnO$_3$ platelets is shown in Fig. 4.  For our
annealing experiments we used a conventional chamber furnace (carbolite CWF~13/13) which allows
for temperatures from room temperature up to 1550~K exceeding considerably the $T_C \approx 1270$~K of
our YMnO$_3$ samples. First we performed a pre-annealing at 1420~K for 24~h under constant oxygen
flow of 0.2~l/min. In the second step the samples were annealed again with a different cooling
rate. Up to 8~K/min the temperature gradient was controlled by the furnace, higher
rates were obtained by removing the fused-silica cell with the sample from the furnace and
measured by an infrared camera (InfraTec VarioCAM). The annealing was followed by thinning
the samples down by 5-10~$\mu$m using Al$_2$O$_3$ and 
polishing chemically-mechanically with 32~nm grain-size slurry. In the last
step the ferroelectric domain structure was imaged using piezoresonse force microscopy
\cite{Jungk_et_al:2010,Chae_et_al:2012}. We used a
commercial scanning force microscope (Solaris, NT-MDT) operated in contact mode and applied an
AC-voltage of 14\,V$_{\rm pp}$ at a frequency of $\sim$40~kHz to a conductive Pt-Ir coated
probe (NSC~35, Mikromasch). The out-of-plane component of the piezoelectric response was recorded
by the in-phase output channel of an external lock-in amplifier (SR830, Stanford Research) with a
typical sensitivity of 200~$\mu$V and time constant of 10~ms. 
Finally, the area density of vortices was extracted from the PFM images. 

\begin{figure}
\centering
\includegraphics[width=0.75\linewidth]{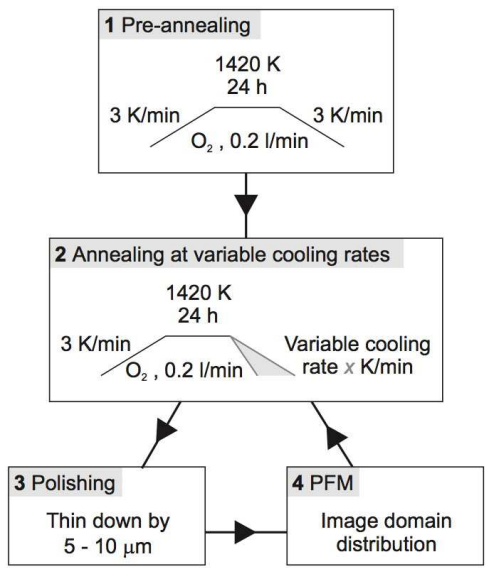}
\caption{
Sketch of the annealing procedure. After pre-annealing the samples are heated to 1420~K
with a hold time of 24~h and subsequent cooling at rates between 0.1 and 1360~K/min. After the
annealing cycle they are thinned down 5-10~$\mu$m by polishing in order to access the domain
structure formed in the true three-dimensional bulk of the sample.}
\label{Supp1}
\end{figure}

\begin{figure*}
 \centering
 \includegraphics[width=0.8\linewidth]{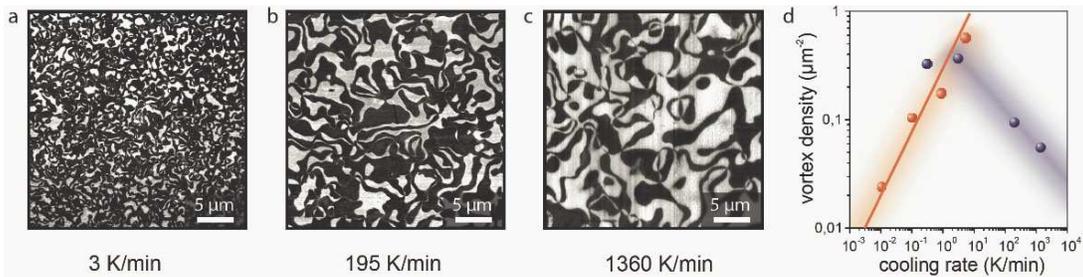}
 \caption{(a) to (c) 
Distribution of ferroelectric domains in $z$-oriented YMnO$_3$ samples after annealing
cycles at different cooling rates. The images are obtained using PFM on an area of
$30\times 30$~$\mu\text{m}^2$ and reveal a striking `anti-Kibble-Zurek'' behavior with higher
cooling rates leading to larger domains. (d) Areal vortex core density as a function of cooling 
rate for slow cooling (red circles, Ref.\cite{Chae_et_al:2012}) and fast 
cooling (blue circles, this work). 
Note the turnover in the cooling-rate dependence of the vortex-core
density occurring between rates of 0.3 and 3~K/min.
The red solid line is the result of our {\it ab initio} application of the Kibble-Zurek scenario
with parameters from first-principles calculations.
}
 \label{Exp}
\end{figure*}

In Fig.\ 5 (a-c) we show the measured distribution of domains in YMnO$_3$ for fast cooling 
rates of 3, 195, and 1360~K/min, and in Fig. 5(d) we plot our measured vortex core densities
as a function of cooling rate (blue circles). The independence of the vortex core density to 
the choice of $R$ is confirmed by the smooth connection between the ErMnO$_3$ (red circles)
and YMnO$_3$ (blue circles)
data points in Fig.\ 5. 
Following a crossover
occurring between 0.3 and 3~K/min (see Fig. 5(d)) we observe a striking
``anti-Kibble-Zurek'' behavior in the measured density of domain vortex cores as a function
of cooling rate: 
Within the fast-cooling regime, an increase in the cooling rate leads to a {\it lowering} of the
density of vortex cores (larger domains); this behaviour is opposite to that predicted by the 
standard Kibble-Zurek
mechanism. We emphasize that the behaviour is highly reproducible: We repeated our measurements on
YMnO$_3$ samples grown in different batches, and verified that there are no drift effects in
consecutive annealing cycles. 
We therefore conclude that our results 
indicate an evolution out of the Kibble-Zurek regime at a cooling rate of $\sim$ 1~K/min
in the hexagonal manganites. This crossover point corresponds to a correlation length (and
hence a crossover domain size) of $\sim 2.2$~$\mu$m, and a relaxation time 
of $\sim 4.1 \times 10^{-4}$~s, with a characteristic information transfer
velocity of $\sim 5.4 \times 10^{-3}$ m/s, considerably reduced from the speed of sound
by the critical slowing down.

\section{Discussion: Possible origins of the Kibble-Zurek / ``anti-Kibble-Zurek'' crossover}

A number of possible deviations from Kibble-Zurek behavior have been discussed in the
literature, but none of them are consistent with our measurements.
Zurek\cite{Yates/Zurek:1998} showed that vortex--anti-vortex annihilation becomes significant at
fast quench rates where domains are smaller and topological defects are closer together.
Such vortex--anti-vortex annihiliation causes a leveling off of the rate 
at which the density of vortex cores increases with cooling rate, but not the decrease 
in density that we observe.
The effect of nonlinearity in the quench rate on the density of defects has been
calculated to yield a modified scaling law\cite{Mondal/Sengupta/Sen:2009}, 
which again would not cause our observed turn-around at fast cooling rates.
In Ref.\cite{Chae_et_al:2012} it
was suggested that the observed production of defect--anti-defect pairs could be the result of a
Kosterlitz-Thouless transition\cite{Kosterlitz/Thouless:1973}, in which 
vortex--anti-vortex pairs are formed above the transition temperature and annihilate as the system
is cooled. As a result, more vortex cores survive during a fast quench when the pairs do not have
time to annihilate. This is the opposite of our observed fast-quenching behavior. In addition, 
a Kosterlitz-Thouless system would show a dramatic change in the density of vortices after repeated 
annealing cycles as well
as a dependence on the temperature at which the quench begins, neither of which we observe.

A possible extrinsic influence on the domain structure could be differences in chemical
defect concentration caused by the different cooling rates, such as off-stoichiometry, anti-site
formation, or charge screening at the domain-walls. To test for this possibility, we heated our
samples to within 2\% of the the transition temperature, and annealed them at this temperature for
six hours under the conditions described above. No changes in the domain structure were observed
on heating until 1270~K (just below $T_C$), at which point minor isolated domain wall movements, 
and, once, the
formation of a vortex--anti-vortex pair were observed. On heating to 1320~K (just above $T_C$), 
a completely new, but
statistically identical domain pattern was obtained. These data and the aforementioned
reproducibility of our data points in Fig.\ 5 show that chemical drift effects do not play
a role.

We conclude therefore, that the most likely origin of the transition out of the Kibble-Zurek
regime in $R$MnO$_3$ is the breakdown of the assumption that the system is able to respond
adiabatically to the cooling at high temperatures, above the freeze-out temperature 
$T_C + \Delta T_f$. 
Our observed behavior in the fast-cooling regime -- slower cooling leading to a larger number of
smaller domains -- is of course reminiscent of nucleation-dominated behavior, with an activation
energy for formation of the low symmetry phase from the high symmetry phase. Nucleation-dominated
phase transitions show charactistic first-order behavior, and a longer time 
spent at the transition allows a larger number of smaller domains to nucleate. 
We note that at the freeze-out temperature corresponding to the crossover quench-rate, the 
order parameter for the trimerization, 
$ \eta = \left( \frac{T-T_{C}}{T_{C}} \right)^\beta$, has already reached 0.5\% of its
saturation value, taking the experimental value of $\beta = 0.29$\cite{Gibbs_et_al:2011}. 
It is possible that this provides a sufficient discontinuity from the zero value to induce 
a first-order response.
An alternative scenario is the fluctuation-induced first-order behaviour proposed 
for prototypical second-order phase
transitions such as the normal-to-superconducting transition and for the nematic-smectic 
transition in liquid crystals\cite{Halperin_et_al:1974}, both of which belong to the same
universality class -- the 3D XY model -- as the hexagonal manganites. 
Such an induced-first-order transition could also explain the current controversy regarding the 
order of the trimerization transition in the hexagonal manganites, with most experiments 
showing second-order behavior, but occasional reports of first-order characteristics.

\section{Summary}

In Summary, we have shown that the multiferroic hexagonal manganites, $R$MnO$_3$, are model
systems for testing the Kibble-Zurek scenario.
Mathematically, they fulfil the symmetry requirements for the formation of topological defects,
and practically, the defects are readily detectable, the quench rate can be varied over a wide
range of relevant timescales, and extrinsic factors that might influence the phase transition
behavior are absent. Our quantitative calculations of topological defect density as a function of
cooling rate using the conventional Kibble-Zurek model and parameters obtained using density
functional theory, yield excellent agreeement with literature data in the slow cooling limit where
the conventional Kibble-Zurek mechanism is applicable. Our measurements of defect density at
fast cooling rates, however, reveal a surprising, apparently ``anti-Kibble-Zurek'' behavior
in which faster cooling yields lower defect densities, reminiscent of a nucleation-dominated phase 
transition. 
Since the expansion of the universe during the inflationary period
was rapid\cite{Guth:1981, Starobinsky:1982}, the turn-around in vortex density which we
find here at fast-cooling rates may be applicable, and 
the observed deficiency of cosmic strings in the cosmic microwave
background\cite{Hindmarsh/Ringeval/Suyama:2010} could be rationalised by the decrease in defect density observed
in the anti-Kibble-Zurek regime.

\section*{Acknowledgments}

We thank Pierre Tol\'edano, Maxim Mostovoy and Ali Mozaffari for invaluable discussions,
Sang-Wook Cheong for sharing the data in Figure 5(d) prior to its publication, and the Cosmic
Superstrings Meeting (Portsmouth, UK, 2011) for allowing us to attend and discuss our ideas. 
This work was supported financially by the ETH Z\"{u}rich. 
MF thanks the IMI Program of the National Science Foundation under Award No.\ DMR-0843934, 
managed by the International Center for Materials Research, UC Santa Barbara, for sabbatical
support.

\bibliography{Manuscript}

\begin{thebibliography}{49}
\expandafter\ifx\csname natexlab\endcsname\relax\def\natexlab#1{#1}\fi
\expandafter\ifx\csname bibnamefont\endcsname\relax
  \def\bibnamefont#1{#1}\fi
\expandafter\ifx\csname bibfnamefont\endcsname\relax
  \def\bibfnamefont#1{#1}\fi
\expandafter\ifx\csname citenamefont\endcsname\relax
  \def\citenamefont#1{#1}\fi
\expandafter\ifx\csname url\endcsname\relax
  \def\url#1{\texttt{#1}}\fi
\expandafter\ifx\csname urlprefix\endcsname\relax\def\urlprefix{URL }\fi
\providecommand{\bibinfo}[2]{#2}
\providecommand{\eprint}[2][]{\url{#2}}

\bibitem[{\citenamefont{Chae et~al.}(2012)\citenamefont{Chae, Lee, Horibe,
  Tanimura, Mori, Gao, Carr, and Cheong}}]{Chae_et_al:2012}
\bibinfo{author}{\bibfnamefont{S.~C.} \bibnamefont{Chae}},
  \bibinfo{author}{\bibfnamefont{N.}~\bibnamefont{Lee}},
  \bibinfo{author}{\bibfnamefont{Y.}~\bibnamefont{Horibe}},
  \bibinfo{author}{\bibfnamefont{M.}~\bibnamefont{Tanimura}},
  \bibinfo{author}{\bibfnamefont{S.}~\bibnamefont{Mori}},
  \bibinfo{author}{\bibfnamefont{B.}~\bibnamefont{Gao}},
  \bibinfo{author}{\bibfnamefont{S.}~\bibnamefont{Carr}}, \bibnamefont{and}
  \bibinfo{author}{\bibfnamefont{S.-W.} \bibnamefont{Cheong}},
  \bibinfo{journal}{Phys. Rev. Lett., in press}  (\bibinfo{year}{2012}),
  \eprint{arXiv:1203.5371v1}.

\bibitem[{\citenamefont{Yates and Zurek}(1998)}]{Yates/Zurek:1998}
\bibinfo{author}{\bibfnamefont{A.}~\bibnamefont{Yates}} \bibnamefont{and}
  \bibinfo{author}{\bibfnamefont{W.~H.} \bibnamefont{Zurek}},
  \bibinfo{journal}{Phys. Rev. Lett.} \textbf{\bibinfo{volume}{80}},
  \bibinfo{pages}{5477} (\bibinfo{year}{1998}).

\bibitem[{\citenamefont{Hindmarsh et~al.}(2010)\citenamefont{Hindmarsh,
  Ringeval, and Suyama}}]{Hindmarsh/Ringeval/Suyama:2010}
\bibinfo{author}{\bibfnamefont{M.}~\bibnamefont{Hindmarsh}},
  \bibinfo{author}{\bibfnamefont{C.}~\bibnamefont{Ringeval}}, \bibnamefont{and}
  \bibinfo{author}{\bibfnamefont{T.}~\bibnamefont{Suyama}},
  \bibinfo{journal}{Phys. Rev. D} \textbf{\bibinfo{volume}{D81}},
  \bibinfo{pages}{063505} (\bibinfo{year}{2010}).

\bibitem[{\citenamefont{Vilenkin and Shellard}(1994)}]{CosmicStrings}
\bibinfo{author}{\bibfnamefont{A.}~\bibnamefont{Vilenkin}} \bibnamefont{and}
  \bibinfo{author}{\bibfnamefont{E.~P.~S.} \bibnamefont{Shellard}},
  \emph{\bibinfo{title}{Cosmic Strings and Other Topological Defects}}
  (\bibinfo{publisher}{Cambridge University Press}, \bibinfo{year}{1994}).

\bibitem[{\citenamefont{Vachaspati}(2009)}]{Vachaspati:2009}
\bibinfo{author}{\bibfnamefont{T.}~\bibnamefont{Vachaspati}},
  \bibinfo{journal}{Phys. Rev. D} \textbf{\bibinfo{volume}{80}},
  \bibinfo{pages}{063502} (\bibinfo{year}{2009}).

\bibitem[{\citenamefont{Bevis et~al.}(2008)\citenamefont{Bevis, Hindmarsh,
  Kunz, and Urrestilla}}]{Bevis/Hindmarsh/Kunz/Urrestilla:2008}
\bibinfo{author}{\bibfnamefont{N.}~\bibnamefont{Bevis}},
  \bibinfo{author}{\bibfnamefont{M.}~\bibnamefont{Hindmarsh}},
  \bibinfo{author}{\bibfnamefont{M.}~\bibnamefont{Kunz}}, \bibnamefont{and}
  \bibinfo{author}{\bibfnamefont{J.}~\bibnamefont{Urrestilla}},
  \bibinfo{journal}{Phys. Rev. Lett.} \textbf{\bibinfo{volume}{100}},
  \bibinfo{pages}{021301} (\bibinfo{year}{2008}).

\bibitem[{\citenamefont{Jeannerot et~al.}(2003)\citenamefont{Jeannerot, Rocher,
  and Sakellariadou}}]{Jeannerot/Rocher/Sakellariadou:2003}
\bibinfo{author}{\bibfnamefont{R.}~\bibnamefont{Jeannerot}},
  \bibinfo{author}{\bibfnamefont{J.}~\bibnamefont{Rocher}}, \bibnamefont{and}
  \bibinfo{author}{\bibfnamefont{M.}~\bibnamefont{Sakellariadou}},
  \bibinfo{journal}{Phys. Rev. D} \textbf{\bibinfo{volume}{68}},
  \bibinfo{pages}{103514} (\bibinfo{year}{2003}).

\bibitem[{\citenamefont{Kibble}(1976)}]{Kibble:1976}
\bibinfo{author}{\bibfnamefont{T.~W.~B.} \bibnamefont{Kibble}},
  \bibinfo{journal}{J. Phys.} \textbf{\bibinfo{volume}{A9}},
  \bibinfo{pages}{1387} (\bibinfo{year}{1976}).

\bibitem[{\citenamefont{Zurek}(1985)}]{Zurek:1985}
\bibinfo{author}{\bibfnamefont{W.~H.} \bibnamefont{Zurek}},
  \bibinfo{journal}{Nature} \textbf{\bibinfo{volume}{317}},
  \bibinfo{pages}{505} (\bibinfo{year}{1985}).

\bibitem[{\citenamefont{Hendry et~al.}(1994)\citenamefont{Hendry, Lawson, Lee,
  McClintock, and Williams}}]{Hendry:1994}
\bibinfo{author}{\bibfnamefont{P.~C.} \bibnamefont{Hendry}},
  \bibinfo{author}{\bibfnamefont{N.~S.} \bibnamefont{Lawson}},
  \bibinfo{author}{\bibfnamefont{R.~A.~M.} \bibnamefont{Lee}},
  \bibinfo{author}{\bibfnamefont{P.~V.~E.} \bibnamefont{McClintock}},
  \bibnamefont{and} \bibinfo{author}{\bibfnamefont{C.~D.~H.}
  \bibnamefont{Williams}}, \bibinfo{journal}{Nature}
  \textbf{\bibinfo{volume}{368}}, \bibinfo{pages}{315} (\bibinfo{year}{1994}).

\bibitem[{\citenamefont{Dodd et~al.}(1998)\citenamefont{Dodd, Hendry, Lawson,
  McClintock, and Williams}}]{Dodd:1998}
\bibinfo{author}{\bibfnamefont{M.~E.} \bibnamefont{Dodd}},
  \bibinfo{author}{\bibfnamefont{P.~C.} \bibnamefont{Hendry}},
  \bibinfo{author}{\bibfnamefont{N.~S.} \bibnamefont{Lawson}},
  \bibinfo{author}{\bibfnamefont{P.~V.~E.} \bibnamefont{McClintock}},
  \bibnamefont{and} \bibinfo{author}{\bibfnamefont{C.~D.~H.}
  \bibnamefont{Williams}}, \bibinfo{journal}{Phys. Rev. Lett.}
  \textbf{\bibinfo{volume}{81}}, \bibinfo{pages}{3703} (\bibinfo{year}{1998}).

\bibitem[{\citenamefont{Rivers}(2000)}]{Rivers:2000}
\bibinfo{author}{\bibfnamefont{R.~J.} \bibnamefont{Rivers}},
  \bibinfo{journal}{Phys. Rev. Lett.} \textbf{\bibinfo{volume}{84}},
  \bibinfo{pages}{1248} (\bibinfo{year}{2000}).

\bibitem[{\citenamefont{Bauerle et~al.}(1996)\citenamefont{Bauerle, Bunkov,
  Fisher, Godfrin, and Pickett}}]{Bauerle_et_al:1996}
\bibinfo{author}{\bibfnamefont{C.}~\bibnamefont{Bauerle}},
  \bibinfo{author}{\bibfnamefont{Y.~M.} \bibnamefont{Bunkov}},
  \bibinfo{author}{\bibfnamefont{S.~N.} \bibnamefont{Fisher}},
  \bibinfo{author}{\bibfnamefont{H.}~\bibnamefont{Godfrin}}, \bibnamefont{and}
  \bibinfo{author}{\bibfnamefont{G.~R.} \bibnamefont{Pickett}},
  \bibinfo{journal}{Nature} \textbf{\bibinfo{volume}{382}},
  \bibinfo{pages}{332} (\bibinfo{year}{1996}).

\bibitem[{\citenamefont{Ruutu et~al.}(1996)\citenamefont{Ruutu, Eltsov, Gill,
  Kibble, Krusius, Makhlin, and Xu}}]{Ruutu:1996}
\bibinfo{author}{\bibfnamefont{V.~M.~H.} \bibnamefont{Ruutu}},
  \bibinfo{author}{\bibfnamefont{V.~B.} \bibnamefont{Eltsov}},
  \bibinfo{author}{\bibfnamefont{A.~J.} \bibnamefont{Gill}},
  \bibinfo{author}{\bibfnamefont{T.~W.~B.} \bibnamefont{Kibble}},
  \bibinfo{author}{\bibfnamefont{M.}~\bibnamefont{Krusius}},
  \bibinfo{author}{\bibfnamefont{B.~V. G.~E.} \bibnamefont{Makhlin},
  \bibfnamefont{Yu. G. annd~Placais}}, \bibnamefont{and}
  \bibinfo{author}{\bibfnamefont{W.}~\bibnamefont{Xu}},
  \bibinfo{journal}{Nature} \textbf{\bibinfo{volume}{382}},
  \bibinfo{pages}{334} (\bibinfo{year}{1996}).

\bibitem[{\citenamefont{Monaco et~al.}(2009)\citenamefont{Monaco, Mygind,
  Rivers, and Koshelets}}]{Monaco_et_al:2009}
\bibinfo{author}{\bibfnamefont{R.}~\bibnamefont{Monaco}},
  \bibinfo{author}{\bibfnamefont{J.}~\bibnamefont{Mygind}},
  \bibinfo{author}{\bibfnamefont{R.~J.} \bibnamefont{Rivers}},
  \bibnamefont{and} \bibinfo{author}{\bibfnamefont{V.~P.}
  \bibnamefont{Koshelets}}, \bibinfo{journal}{Phys. Rev. B}
  \textbf{\bibinfo{volume}{80}}, \bibinfo{pages}{180501}
  (\bibinfo{year}{2009}).

\bibitem[{\citenamefont{Carmi and Polturak}(1999)}]{Carmi/Polturak:1999}
\bibinfo{author}{\bibfnamefont{R.}~\bibnamefont{Carmi}} \bibnamefont{and}
  \bibinfo{author}{\bibfnamefont{E.}~\bibnamefont{Polturak}},
  \bibinfo{journal}{Phys. Rev. B} \textbf{\bibinfo{volume}{60}},
  \bibinfo{pages}{7595} (\bibinfo{year}{1999}).

\bibitem[{\citenamefont{Weiler et~al.}(2008)\citenamefont{Weiler, Neely,
  Scherer, Bradley, Davis, and Anderson}}]{Weiler_et_al:2008}
\bibinfo{author}{\bibfnamefont{C.~N.} \bibnamefont{Weiler}},
  \bibinfo{author}{\bibfnamefont{T.~W.} \bibnamefont{Neely}},
  \bibinfo{author}{\bibfnamefont{D.~R.} \bibnamefont{Scherer}},
  \bibinfo{author}{\bibfnamefont{S.}~\bibnamefont{Bradley},
  \bibfnamefont{Ashton}},
  \bibinfo{author}{\bibfnamefont{J.}~\bibnamefont{Davis},
  \bibfnamefont{Matthew}}, \bibnamefont{and}
  \bibinfo{author}{\bibfnamefont{B.~P.} \bibnamefont{Anderson}},
  \bibinfo{journal}{Nature} \textbf{\bibinfo{volume}{455}},
  \bibinfo{pages}{948} (\bibinfo{year}{2008}).

\bibitem[{\citenamefont{Ducci et~al.}(1999)\citenamefont{Ducci, Ramazza,
  Gonzalez-Vinas, and Arecchi}}]{Ducci_et_al:1999}
\bibinfo{author}{\bibfnamefont{S.}~\bibnamefont{Ducci}},
  \bibinfo{author}{\bibfnamefont{P.~L.} \bibnamefont{Ramazza}},
  \bibinfo{author}{\bibfnamefont{W.}~\bibnamefont{Gonzalez-Vinas}},
  \bibnamefont{and} \bibinfo{author}{\bibfnamefont{F.~T.}
  \bibnamefont{Arecchi}}, \bibinfo{journal}{Phys. Rev. Lett.}
  \textbf{\bibinfo{volume}{83}}, \bibinfo{pages}{5210} (\bibinfo{year}{1999}).

\bibitem[{\citenamefont{Chuang et~al.}(1991)\citenamefont{Chuang, Durrer,
  Turok, and Yurke}}]{Chuang_et_al:1991}
\bibinfo{author}{\bibfnamefont{I.}~\bibnamefont{Chuang}},
  \bibinfo{author}{\bibfnamefont{R.}~\bibnamefont{Durrer}},
  \bibinfo{author}{\bibfnamefont{N.}~\bibnamefont{Turok}}, \bibnamefont{and}
  \bibinfo{author}{\bibfnamefont{B.}~\bibnamefont{Yurke}},
  \bibinfo{journal}{Science} \textbf{\bibinfo{volume}{251}},
  \bibinfo{pages}{1336} (\bibinfo{year}{1991}).

\bibitem[{\citenamefont{van Aken et~al.}(2004)\citenamefont{van Aken, Palstra,
  Filippetti, and Spaldin}}]{vanAken_et_al:2004}
\bibinfo{author}{\bibfnamefont{B.~B.} \bibnamefont{van Aken}},
  \bibinfo{author}{\bibfnamefont{T.~T.~M.} \bibnamefont{Palstra}},
  \bibinfo{author}{\bibfnamefont{A.}~\bibnamefont{Filippetti}},
  \bibnamefont{and} \bibinfo{author}{\bibfnamefont{N.~A.}
  \bibnamefont{Spaldin}}, \bibinfo{journal}{Nature Mater.}
  \textbf{\bibinfo{volume}{3}}, \bibinfo{pages}{164} (\bibinfo{year}{2004}).

\bibitem[{\citenamefont{Fennie and Rabe}(2005)}]{Fennie/Rabe_YMO:2005}
\bibinfo{author}{\bibfnamefont{C.~J.} \bibnamefont{Fennie}} \bibnamefont{and}
  \bibinfo{author}{\bibfnamefont{K.~M.} \bibnamefont{Rabe}},
  \bibinfo{journal}{Phys. Rev. B} \textbf{\bibinfo{volume}{72}},
  \bibinfo{pages}{100103(R)} (\bibinfo{year}{2005}).

\bibitem[{\citenamefont{Fiebig et~al.}(2002)\citenamefont{Fiebig, Lottermoser,
  Fr{\"o}hlich, Goltsev, and Pisarev}}]{Fiebig_et_al:2002}
\bibinfo{author}{\bibfnamefont{M.}~\bibnamefont{Fiebig}},
  \bibinfo{author}{\bibfnamefont{T.}~\bibnamefont{Lottermoser}},
  \bibinfo{author}{\bibfnamefont{D.}~\bibnamefont{Fr{\"o}hlich}},
  \bibinfo{author}{\bibfnamefont{A.~V.} \bibnamefont{Goltsev}},
  \bibnamefont{and} \bibinfo{author}{\bibfnamefont{R.~V.}
  \bibnamefont{Pisarev}}, \bibinfo{journal}{Nature}
  \textbf{\bibinfo{volume}{419}}, \bibinfo{pages}{818} (\bibinfo{year}{2002}).

\bibitem[{\citenamefont{Lottermoser et~al.}(2004)\citenamefont{Lottermoser,
  Lonkai, Amann, Hohlwein, Ihringer, and Fiebig}}]{Lottermoser_et_al:2004}
\bibinfo{author}{\bibfnamefont{T.}~\bibnamefont{Lottermoser}},
  \bibinfo{author}{\bibfnamefont{T.}~\bibnamefont{Lonkai}},
  \bibinfo{author}{\bibfnamefont{U.}~\bibnamefont{Amann}},
  \bibinfo{author}{\bibfnamefont{D.}~\bibnamefont{Hohlwein}},
  \bibinfo{author}{\bibfnamefont{J.}~\bibnamefont{Ihringer}}, \bibnamefont{and}
  \bibinfo{author}{\bibfnamefont{M.}~\bibnamefont{Fiebig}},
  \bibinfo{journal}{Nature} \textbf{\bibinfo{volume}{430}},
  \bibinfo{pages}{541} (\bibinfo{year}{2004}).

\bibitem[{\citenamefont{Choi et~al.}(2010)\citenamefont{Choi, Horibe, Yi, Choi,
  Wu, and Cheong}}]{Choi_et_al:2010}
\bibinfo{author}{\bibfnamefont{T.}~\bibnamefont{Choi}},
  \bibinfo{author}{\bibfnamefont{Y.}~\bibnamefont{Horibe}},
  \bibinfo{author}{\bibfnamefont{H.~T.} \bibnamefont{Yi}},
  \bibinfo{author}{\bibfnamefont{Y.~J.} \bibnamefont{Choi}},
  \bibinfo{author}{\bibfnamefont{W.}~\bibnamefont{Wu}}, \bibnamefont{and}
  \bibinfo{author}{\bibfnamefont{S.-W.} \bibnamefont{Cheong}},
  \bibinfo{journal}{Nat. Mater.} \textbf{\bibinfo{volume}{9}},
  \bibinfo{pages}{253} (\bibinfo{year}{2010}).

\bibitem[{\citenamefont{Meier et~al.}(2012)\citenamefont{Meier, Seidel, Cano,
  Delaney, Kumagai, Mostovoy, Spaldin, Ramesh, and Fiebig}}]{Meier_et_al:2012}
\bibinfo{author}{\bibfnamefont{D.}~\bibnamefont{Meier}},
  \bibinfo{author}{\bibfnamefont{J.}~\bibnamefont{Seidel}},
  \bibinfo{author}{\bibfnamefont{A.}~\bibnamefont{Cano}},
  \bibinfo{author}{\bibfnamefont{K.}~\bibnamefont{Delaney}},
  \bibinfo{author}{\bibfnamefont{Y.}~\bibnamefont{Kumagai}},
  \bibinfo{author}{\bibfnamefont{M.}~\bibnamefont{Mostovoy}},
  \bibinfo{author}{\bibfnamefont{N.~A.} \bibnamefont{Spaldin}},
  \bibinfo{author}{\bibfnamefont{R.}~\bibnamefont{Ramesh}}, \bibnamefont{and}
  \bibinfo{author}{\bibfnamefont{M.}~\bibnamefont{Fiebig}},
  \bibinfo{journal}{Nat. Mater.} \textbf{\bibinfo{volume}{11}},
  \bibinfo{pages}{284} (\bibinfo{year}{2012}).

\bibitem[{\citenamefont{Yakel et~al.}(1963)\citenamefont{Yakel, Koehler,
  Bertaut, and Forrat}}]{Yakel_et_al:1963}
\bibinfo{author}{\bibfnamefont{H.~L.} \bibnamefont{Yakel}},
  \bibinfo{author}{\bibfnamefont{W.~C.} \bibnamefont{Koehler}},
  \bibinfo{author}{\bibfnamefont{E.~F.} \bibnamefont{Bertaut}},
  \bibnamefont{and} \bibinfo{author}{\bibfnamefont{E.~F.}
  \bibnamefont{Forrat}}, \bibinfo{journal}{Acta Cryst.}
  \textbf{\bibinfo{volume}{16}}, \bibinfo{pages}{957} (\bibinfo{year}{1963}).

\bibitem[{\citenamefont{Lonkai et~al.}(2004)\citenamefont{Lonkai, Tomuta,
  Amann, Ihringer, Hendrikx, Tobbens, and Mydosh}}]{Lonkai_et_al:2004}
\bibinfo{author}{\bibfnamefont{T.}~\bibnamefont{Lonkai}},
  \bibinfo{author}{\bibfnamefont{D.~G.} \bibnamefont{Tomuta}},
  \bibinfo{author}{\bibfnamefont{U.}~\bibnamefont{Amann}},
  \bibinfo{author}{\bibfnamefont{J.}~\bibnamefont{Ihringer}},
  \bibinfo{author}{\bibfnamefont{R.~W.~A.} \bibnamefont{Hendrikx}},
  \bibinfo{author}{\bibfnamefont{D.~M.} \bibnamefont{Tobbens}},
  \bibnamefont{and} \bibinfo{author}{\bibfnamefont{J.~A.}
  \bibnamefont{Mydosh}}, \bibinfo{journal}{Phys. Rev. B}
  \textbf{\bibinfo{volume}{69}}, \bibinfo{pages}{134108}
  (\bibinfo{year}{2004}).

\bibitem[{\citenamefont{Jungk et~al.}(2010)\citenamefont{Jungk, Hoffmann,
  Fiebig, and Soergel}}]{Jungk_et_al:2010}
\bibinfo{author}{\bibfnamefont{T.}~\bibnamefont{Jungk}},
  \bibinfo{author}{\bibfnamefont{{\'A}.}~\bibnamefont{Hoffmann}},
  \bibinfo{author}{\bibfnamefont{M.}~\bibnamefont{Fiebig}}, \bibnamefont{and}
  \bibinfo{author}{\bibfnamefont{E.}~\bibnamefont{Soergel}},
  \bibinfo{journal}{Appl. Phys. Lett.} \textbf{\bibinfo{volume}{97}},
  \bibinfo{pages}{012904} (\bibinfo{year}{2010}).

\bibitem[{\citenamefont{Artyukhin et~al.}(2012)\citenamefont{Artyukhin,
  Delaney, Spaldin, and Mostovoy}}]{Artyukhin_et_al:2012}
\bibinfo{author}{\bibfnamefont{S.}~\bibnamefont{Artyukhin}},
  \bibinfo{author}{\bibfnamefont{K.~T.} \bibnamefont{Delaney}},
  \bibinfo{author}{\bibfnamefont{N.~A.} \bibnamefont{Spaldin}},
  \bibnamefont{and} \bibinfo{author}{\bibfnamefont{M.}~\bibnamefont{Mostovoy}},
  \bibinfo{journal}{In preparation}  (\bibinfo{year}{2012}).

\bibitem[{\citenamefont{Kervaire and Milnor}(1963)}]{Kervaire:1963}
\bibinfo{author}{\bibfnamefont{M.~A.} \bibnamefont{Kervaire}} \bibnamefont{and}
  \bibinfo{author}{\bibfnamefont{J.~W.} \bibnamefont{Milnor}},
  \bibinfo{journal}{Ann. of Math., Second Series}
  \textbf{\bibinfo{volume}{77}}, \bibinfo{pages}{504} (\bibinfo{year}{1963}).

\bibitem[{\citenamefont{Kibble}(2000)}]{Kibble:2000}
\bibinfo{author}{\bibfnamefont{T.~W.~B.} \bibnamefont{Kibble}},
  \emph{\bibinfo{title}{Classification of topological defects and their
  relevance to cosmology and elsewhere, in Topological Defects and the
  Non-Equilibrium Dynamics of Symmetry Breaking Phase Transitions}}, vol.
  \bibinfo{volume}{C 549} of \emph{\bibinfo{series}{NATO Science Series}}
  (\bibinfo{publisher}{Kluwer Academic Publishers}, \bibinfo{year}{2000}).

\bibitem[{\citenamefont{Hindmarsh and Rajantie}(2000)}]{PhysRevLett.85.4660}
\bibinfo{author}{\bibfnamefont{M.}~\bibnamefont{Hindmarsh}} \bibnamefont{and}
  \bibinfo{author}{\bibfnamefont{A.}~\bibnamefont{Rajantie}},
  \bibinfo{journal}{Phys. Rev. Lett.} \textbf{\bibinfo{volume}{85}},
  \bibinfo{pages}{4660} (\bibinfo{year}{2000}).

\bibitem[{\citenamefont{Kibble}(2003)}]{Kibble:2003}
\bibinfo{author}{\bibfnamefont{T.~W.~B.} \bibnamefont{Kibble}}, in
  \emph{\bibinfo{booktitle}{Patterns of Symmetry Breaking}}
  (\bibinfo{year}{2003}), vol. \bibinfo{volume}{127} of
  \emph{\bibinfo{series}{NATO Science Series II:Mathematics, Physics and
  Chemistry}}, pp. \bibinfo{pages}{3--36}, \bibinfo{note}{proceedings of the
  Conference of the NATO-Advanced-Study-Institute on Patterns of Symmetry
  Breaking, Cracow, Poland, September 2002}.

\bibitem[{\citenamefont{Liechtenstein et~al.}(1995)\citenamefont{Liechtenstein,
  Anisimov, and Zaanen}}]{Lichtenstein_et_al:1995}
\bibinfo{author}{\bibfnamefont{A.~I.} \bibnamefont{Liechtenstein}},
  \bibinfo{author}{\bibfnamefont{V.~I.} \bibnamefont{Anisimov}},
  \bibnamefont{and} \bibinfo{author}{\bibfnamefont{J.}~\bibnamefont{Zaanen}},
  \bibinfo{journal}{Phys. Rev. B} \textbf{\bibinfo{volume}{52}},
  \bibinfo{pages}{R5467} (\bibinfo{year}{1995}).

\bibitem[{\citenamefont{Bl{\"o}chl}(1994)}]{Bloechl:1994}
\bibinfo{author}{\bibfnamefont{P.~E.} \bibnamefont{Bl{\"o}chl}},
  \bibinfo{journal}{Phys. Rev. B} \textbf{\bibinfo{volume}{50}},
  \bibinfo{pages}{17953} (\bibinfo{year}{1994}).

\bibitem[{\citenamefont{Kresse and Hafner}(1993)}]{Kresse/Hafner:1993}
\bibinfo{author}{\bibfnamefont{G.}~\bibnamefont{Kresse}} \bibnamefont{and}
  \bibinfo{author}{\bibfnamefont{J.}~\bibnamefont{Hafner}},
  \bibinfo{journal}{Phys. Rev. B} \textbf{\bibinfo{volume}{47}},
  \bibinfo{pages}{558} (\bibinfo{year}{1993}).

\bibitem[{\citenamefont{Kresse and Furthm$\ddot{\rm
  u}$ller}(1996)}]{Kresse/Furthmuller:1996}
\bibinfo{author}{\bibfnamefont{G.}~\bibnamefont{Kresse}} \bibnamefont{and}
  \bibinfo{author}{\bibfnamefont{J.}~\bibnamefont{Furthm$\ddot{\rm u}$ller}},
  \bibinfo{journal}{Phys. Rev. B} \textbf{\bibinfo{volume}{54}},
  \bibinfo{pages}{11169} (\bibinfo{year}{1996}).

\bibitem[{\citenamefont{Zhang et~al.}(2012)\citenamefont{Zhang, Wang, Wei, Yu,
  Gu, Hirata, Chen, Jin, Yao, Wang et~al.}}]{Zhang_et_al:2012}
\bibinfo{author}{\bibfnamefont{Q.~H.} \bibnamefont{Zhang}},
  \bibinfo{author}{\bibfnamefont{L.~J.} \bibnamefont{Wang}},
  \bibinfo{author}{\bibfnamefont{X.~K.} \bibnamefont{Wei}},
  \bibinfo{author}{\bibfnamefont{R.~C.} \bibnamefont{Yu}},
  \bibinfo{author}{\bibfnamefont{L.}~\bibnamefont{Gu}},
  \bibinfo{author}{\bibfnamefont{A.}~\bibnamefont{Hirata}},
  \bibinfo{author}{\bibfnamefont{M.~W.} \bibnamefont{Chen}},
  \bibinfo{author}{\bibfnamefont{C.~Q.} \bibnamefont{Jin}},
  \bibinfo{author}{\bibfnamefont{Y.}~\bibnamefont{Yao}},
  \bibinfo{author}{\bibfnamefont{Y.~G.} \bibnamefont{Wang}},
  \bibnamefont{et~al.}, \bibinfo{journal}{Phys. Rev. B}
  \textbf{\bibinfo{volume}{85}}, \bibinfo{pages}{020102 (R)}
  (\bibinfo{year}{2012}).

\bibitem[{\citenamefont{Gonze et~al.}(2002)\citenamefont{Gonze, Beuken,
  Caracas, Detraux, Fuchs, Rignanese, Sindic, Verstraete, Zerah, Jollet
  et~al.}}]{Gonze_et_al:2002}
\bibinfo{author}{\bibfnamefont{X.}~\bibnamefont{Gonze}},
  \bibinfo{author}{\bibfnamefont{J.-M.} \bibnamefont{Beuken}},
  \bibinfo{author}{\bibfnamefont{R.}~\bibnamefont{Caracas}},
  \bibinfo{author}{\bibfnamefont{F.}~\bibnamefont{Detraux}},
  \bibinfo{author}{\bibfnamefont{M.}~\bibnamefont{Fuchs}},
  \bibinfo{author}{\bibfnamefont{G.-M.} \bibnamefont{Rignanese}},
  \bibinfo{author}{\bibfnamefont{L.}~\bibnamefont{Sindic}},
  \bibinfo{author}{\bibfnamefont{M.}~\bibnamefont{Verstraete}},
  \bibinfo{author}{\bibfnamefont{G.}~\bibnamefont{Zerah}},
  \bibinfo{author}{\bibfnamefont{F.}~\bibnamefont{Jollet}},
  \bibnamefont{et~al.}, \bibinfo{journal}{Comp. Mater. Sci.}
  \textbf{\bibinfo{volume}{25}}, \bibinfo{pages}{478} (\bibinfo{year}{2002}).

\bibitem[{\citenamefont{Gonze et~al.}(2009)\citenamefont{Gonze, Amadon,
  Anglade, Beuken, Bottin, Boulanger, Bruneval, Caliste, Caracas, Cote
  et~al.}}]{Gonze_et_al:2009}
\bibinfo{author}{\bibfnamefont{X.}~\bibnamefont{Gonze}},
  \bibinfo{author}{\bibfnamefont{B.}~\bibnamefont{Amadon}},
  \bibinfo{author}{\bibfnamefont{P.-M.} \bibnamefont{Anglade}},
  \bibinfo{author}{\bibfnamefont{J.-M.} \bibnamefont{Beuken}},
  \bibinfo{author}{\bibfnamefont{F.}~\bibnamefont{Bottin}},
  \bibinfo{author}{\bibfnamefont{P.}~\bibnamefont{Boulanger}},
  \bibinfo{author}{\bibfnamefont{F.}~\bibnamefont{Bruneval}},
  \bibinfo{author}{\bibfnamefont{D.}~\bibnamefont{Caliste}},
  \bibinfo{author}{\bibfnamefont{R.}~\bibnamefont{Caracas}},
  \bibinfo{author}{\bibfnamefont{M.}~\bibnamefont{Cote}}, \bibnamefont{et~al.},
  \bibinfo{journal}{Comp. Phys. Commun.} \textbf{\bibinfo{volume}{180}},
  \bibinfo{pages}{2582} (\bibinfo{year}{2009}).

\bibitem[{\citenamefont{Togo et~al.}(2008)\citenamefont{Togo, Oba, and
  Tanaka}}]{phonopy}
\bibinfo{author}{\bibfnamefont{A.}~\bibnamefont{Togo}},
  \bibinfo{author}{\bibfnamefont{F.}~\bibnamefont{Oba}}, \bibnamefont{and}
  \bibinfo{author}{\bibfnamefont{I.}~\bibnamefont{Tanaka}},
  \bibinfo{journal}{Phys. Rev. B} \textbf{\bibinfo{volume}{78}},
  \bibinfo{pages}{134106} (\bibinfo{year}{2008}).

\bibitem[{\citenamefont{Parlinski et~al.}(1997)\citenamefont{Parlinski, Li, and
  Kawazoe}}]{Parlinski_et_al:1997}
\bibinfo{author}{\bibfnamefont{K.}~\bibnamefont{Parlinski}},
  \bibinfo{author}{\bibfnamefont{Z.~Q.} \bibnamefont{Li}}, \bibnamefont{and}
  \bibinfo{author}{\bibfnamefont{Y.}~\bibnamefont{Kawazoe}},
  \bibinfo{journal}{Phys. Rev. Lett.} \textbf{\bibinfo{volume}{78}},
  \bibinfo{pages}{4063} (\bibinfo{year}{1997}).

\bibitem[{\citenamefont{Campostrini et~al.}(2006)\citenamefont{Campostrini,
  Hasenbusch, Pelissetto, and Vicari}}]{Campostrini_et_al:2006}
\bibinfo{author}{\bibfnamefont{M.}~\bibnamefont{Campostrini}},
  \bibinfo{author}{\bibfnamefont{M.}~\bibnamefont{Hasenbusch}},
  \bibinfo{author}{\bibfnamefont{A.}~\bibnamefont{Pelissetto}},
  \bibnamefont{and} \bibinfo{author}{\bibfnamefont{E.}~\bibnamefont{Vicari}},
  \bibinfo{journal}{Phys. Rev. B} \textbf{\bibinfo{volume}{74}},
  \bibinfo{pages}{144506} (\bibinfo{year}{2006}).

\bibitem[{\citenamefont{Mondal et~al.}(2009)\citenamefont{Mondal, Sengupta, and
  Sen}}]{Mondal/Sengupta/Sen:2009}
\bibinfo{author}{\bibfnamefont{S.}~\bibnamefont{Mondal}},
  \bibinfo{author}{\bibfnamefont{K.}~\bibnamefont{Sengupta}}, \bibnamefont{and}
  \bibinfo{author}{\bibfnamefont{D.}~\bibnamefont{Sen}},
  \bibinfo{journal}{Phys. Rev. B} \textbf{\bibinfo{volume}{79}},
  \bibinfo{pages}{045128} (\bibinfo{year}{2009}).

\bibitem[{\citenamefont{Kosterlitz and
  Thouless}(1973)}]{Kosterlitz/Thouless:1973}
\bibinfo{author}{\bibfnamefont{J.~M.} \bibnamefont{Kosterlitz}}
  \bibnamefont{and} \bibinfo{author}{\bibfnamefont{D.~J.}
  \bibnamefont{Thouless}}, \bibinfo{journal}{Journal of Physics C Solid State
  Physics} \textbf{\bibinfo{volume}{6}}, \bibinfo{pages}{1181}
  (\bibinfo{year}{1973}).

\bibitem[{\citenamefont{Gibbs et~al.}(2011)\citenamefont{Gibbs, Knight, and
  Lightfoot}}]{Gibbs_et_al:2011}
\bibinfo{author}{\bibfnamefont{A.~S.} \bibnamefont{Gibbs}},
  \bibinfo{author}{\bibfnamefont{K.~S.} \bibnamefont{Knight}},
  \bibnamefont{and}
  \bibinfo{author}{\bibfnamefont{P.}~\bibnamefont{Lightfoot}},
  \bibinfo{journal}{Phys. Rev. B} \textbf{\bibinfo{volume}{83}},
  \bibinfo{pages}{094111} (\bibinfo{year}{2011}).

\bibitem[{\citenamefont{Halperin et~al.}(1974)\citenamefont{Halperin, Lubensky,
  and Ma}}]{Halperin_et_al:1974}
\bibinfo{author}{\bibfnamefont{B.~I.} \bibnamefont{Halperin}},
  \bibinfo{author}{\bibfnamefont{T.~C.} \bibnamefont{Lubensky}},
  \bibnamefont{and} \bibinfo{author}{\bibfnamefont{S.-K.} \bibnamefont{Ma}},
  \bibinfo{journal}{Phys. Rev. Lett.} \textbf{\bibinfo{volume}{32}},
  \bibinfo{pages}{292} (\bibinfo{year}{1974}).

\bibitem[{\citenamefont{Guth}(1981)}]{Guth:1981}
\bibinfo{author}{\bibfnamefont{A.~H.} \bibnamefont{Guth}},
  \bibinfo{journal}{Phys. Rev. D} \textbf{\bibinfo{volume}{23}},
  \bibinfo{pages}{347} (\bibinfo{year}{1981}).

\bibitem[{\citenamefont{Starobinsky}(1982)}]{Starobinsky:1982}
\bibinfo{author}{\bibfnamefont{A.~A.} \bibnamefont{Starobinsky}},
  \bibinfo{journal}{Physics Letters B} \textbf{\bibinfo{volume}{117}},
  \bibinfo{pages}{175} (\bibinfo{year}{1982}).

\end{thebibliography}

\end{document}